%% file: main-ieee.tex
\documentclass[conference,10pt]{IEEEtran}

\IEEEoverridecommandlockouts
\ifCLASSOPTIONcompsoc
  \usepackage[nocompress]{cite}
\else
  \usepackage{cite}
\fi
\usepackage{amsmath,amssymb,amsfonts}
\usepackage{algorithmic}
\usepackage{graphicx}
\usepackage{textcomp}
\usepackage[dvipsnames]{xcolor}
\usepackage{url}
\usepackage{xspace} % \todo が \xspace を使うため必要
\usepackage{balance}
\usepackage[caption=false,font=normalsize,
   labelfont=sf,textfont=sf]{subfig}
\usepackage[nameinlink]{cleveref}
\usepackage{combelow}

\usepackage{ascmac}
\usepackage{threeparttable}
\usepackage[hang,small,bf]{caption}
\usepackage[subrefformat=parens]{subcaption}

\usepackage{pbalance}

\newboolean{showcomments}
\setboolean{showcomments}{true} % comment this line to deactivate comments
\ifthenelse{\boolean{showcomments}}{
  \newcommand{\nbc}[3]{
    {\colorbox{#3}{\bfseries\sffamily\scriptsize\textcolor{white}{#1}}}%
    {\textcolor{#3}{\textsf\small$\blacktriangleright$\textit{#2}$\blacktriangleleft$}}}
  \newcommand{\todo}[1]{\nbc{TODO}{#1}{blue}\xspace}
}{
  \newcommand{\nbc}[3]{}
  \renewcommand{\todo}[1]{}
}

\usepackage{booktabs}  % 表の横罫 \toprule/\midrule/\bottomrule/\cmidrule（全7表）
\usepackage{tabularx}  % 段幅ちょうどに収める表（3箇所）
\usepackage{multirow}  % 表のセル縦結合（2箇所）

\def\incfig#1#2{\includegraphics[width = #1\linewidth]{#2}}

\newsavebox{\answerbox}

\begin{document}
\title{
GitCF: Reducing Incomplete Changes by Exploiting Multiple Similarities Among Commits
}

\author{
  \IEEEauthorblockN{Yuto Kitabayashi}
  \IEEEauthorblockA{\textit{School of Computing} \\
    \textit{Institute of Science Tokyo}\\
    Yokohama, Japan \\
    y.kitabayashi@sa.comp.isct.ac.jp}
  \and
  \IEEEauthorblockN{Takashi Kobayashi}
  \IEEEauthorblockA{\textit{School of Computing} \\
    \textit{Institute of Science Tokyo}\\
    Yokohama, Japan \\
    ORCID:0000-0002-2235-9992}
}

\maketitle

\begin{abstract}
Software developers often struggle to identify all locations that their changes affect, and incomplete changes frequently result from these omissions.
To mitigate this problem, several approaches have been proposed to recommend additional locations that should be modified together with a developer's current change. A large class of these techniques mines co-change rules from repository revision histories, but because they rely solely on which elements have been modified together, they cannot recommend elements that have rarely co-changed, and they disregard the textual information that accompanies commits. A second class of techniques does exploit textual information, but it derives it from the source code or from a change request rather than from commit history, and it does not use the developer's current change as its input.
To address these limitations, we propose GitCF, a change recommendation method that compares a developer's current change with past changes in the revision history and recommends additional change locations by combining two sources of similarity: the set of modified elements and the textual content of commits, including commit messages, code diffs, and linked issue descriptions.
We evaluate GitCF on 543 incomplete changes from 17 open-source projects, where an incomplete change denotes a commit in which an omitted modification is supplied in a later commit, comparing it against three co-change-rule-based techniques. GitCF outperforms the strongest baseline on all five metrics (MAP, Recall@10, Recall@20, Hit@10, and Hit@20); in particular, it raises Hit@10 from 0.322 to 0.403, and the improvement in average precision is statistically significant.
\end{abstract}


\input{body-e.tex}

\vspace{0.3em}
\noindent\textbf{Acknowledgment: }
This work was partially supported by KAKENHI JP18H03221 and JP22H03567.
During the preparation of this work, the authors used ChatGPT and Claude to improve the readability of the manuscript.
After using these tools, the authors reviewed and edited the content as needed and take full responsibility for the content of the publication.

\bibliographystyle{IEEEtran}
\bibliography{IEEEabrv,ref.bib}

\end{document}

%% file: body-e.tex
%% 
%% To avoid cumbersome term-unification searches, the file is not split too finely.
%% In addition to this file, one file for the method and one for the evaluation are sufficient.
%%

\section{Introduction} \label{sec:intro}
When a modification is applied to a particular part of a software system, its effects may propagate to other locations due to dependencies among program elements (e.g., files, classes, methods). Identifying the locations affected by a change is referred to as change impact analysis. As software evolves, these dependencies can become increasingly complex, making it difficult for developers to anticipate the ripple effects of their modifications. Such challenges may lead to omissions, resulting in incomplete changes.

In this study, we adopt a specific operationalization of this phenomenon. An incomplete change is a commit in which one or more modifications that ought to have been made are omitted, and the omitted modifications are supplied only in a later commit that addresses the resulting problem. This definition is restrictive, because an omission becomes observable only when it is subsequently corrected. Our evaluation therefore covers incomplete changes that satisfy this criterion, and whether the proposed method also assists developers with omissions that are never corrected remains outside the scope of this study.

To reduce the occurrence of incomplete changes, several techniques have been proposed to automatically estimate the ripple effects of a developer's modification. Many existing approaches quantify relationships among program elements using a metric known as \textit{coupling}, recommending elements that exhibit strong relationships with the modified element as potential ripple locations. Representative forms of coupling include Structural Coupling, based on static dependencies such as method calls \cite{briand:icsm1999}; Conceptual Coupling, based on identifier and comment similarity \cite{poshyvanyk:ese2009}; and Evolutionary Coupling, based on co-change relationships in version histories \cite{gall:icsm1998}. Prior studies report that approaches leveraging Conceptual and Evolutionary Coupling can achieve favorable performance \cite{sun:apsi2012}, and Evolutionary Coupling offers properties such as programming-language independence \cite{shen:icsme2021}.

A class of change recommendation techniques based on Evolutionary Coupling applies association rule mining to version histories \cite{zimmermann:icse2004, rolfsnes:saner2016, mori:ipsj2017e}. These techniques extract relationships among program elements that have been modified together and use them as co-change rules to recommend additional change locations. Although co-change–based approaches have shown utility, they rely on sufficiently rich version histories. When a program element has rarely been modified, the number of elements with which it has co-changed is limited, reducing the number of candidates that can be recommended. This limitation is particularly evident in early project stages, when the version history is short.

A second class of techniques addresses a related limitation by incorporating textual information, but it does so by enriching how program elements are represented rather than by enriching the evidence used to rank them. Gethers et al. and Zanjani et al. represent each program element as a sequence of words drawn from its identifiers, comments, and issue descriptions, and then rank elements by the similarity of that representation to a text query. Other work extends this idea to a graph over elements, developers, and tasks. In these approaches, the query is a change request or a declared task, and the similarity is defined between that query and an individual program element. The version history is therefore not the medium through which an unfinished change is located: an element can be surfaced because it is described in similar words, without regard to whether past changes to similar work have actually modified it. When the query is a change request, the approach additionally presupposes that the change has already been filed and described, which is not the situation in which an incomplete change arises.

To address these limitations, this study examines the use of contextual information surrounding a change. 
We propose a change recommendation method that leverages similarity between changes, enabling the use of contextual information such as changed elements and textual descriptions. 
The proposed method extracts features from commits, and provides a mechanism for measuring similarity among commits based on these features.

Although various elements can characterize a change, this study focuses on two categories: changed elements and textual information.
Changed elements represent the program elements modified in a commit, corresponding to the use of Evolutionary Coupling in existing techniques. 
We extend this idea by introducing a measure that reflects the importance of program elements appearing in a commit.
Textual information refers to natural-language content associated with a commit, such as commit messages. 
This study considers three types of textual information: commit messages, issues linked to commits, and diffs from previous commits.

The proposed method retrieves, from the revision history, commits similar to the change a developer is currently performing. 
It then weights the program elements modified in the retrieved commits according to their similarity and ranks them to infer locations that may require modification. Because recommendations are generated independently for each attribute, the results are integrated to produce the final set of recommended change locations. 
This design enables a characterization of change features that incorporates multiple types of information.

We evaluate the proposed method using a dataset of incomplete changes collected from OSS development\cite{yishida:mthesis2021}. 
The results indicate that the proposed method yields higher recommendation accuracy than existing co-change-rule-based techniques on this dataset. 
In addition, the analysis suggests that the method maintains effectiveness even when the version history is short.

The contributions of this work are as follows:
\begin{itemize}
  \item Methods for computing similarity between changes are proposed, one based on changed elements and one based on textual information.
  \item A change recommendation method is investigated from three perspectives: recommendation based on changed elements, recommendation based on textual information, and integration of the two recommendation results.
  \item The method is evaluated using a dataset of incomplete changes that occurred in OSS development\cite{yishida:mthesis2021}, and its accuracy is compared with that of existing co-change-rule-based techniques.
  \item An analysis is conducted to examine the method's behavior when the version history is short.
\end{itemize}

The remainder of this paper is organized as follows. Section \ref{sec:motivating-example} presents a motivating example. Section \ref{sec:relwork} reviews related work. Section \ref{sec:method} describes the proposed method. Section \ref{sec:design} explains the experimental setup, including the research questions and dataset. Section \ref{sec:result} presents and discusses the experimental results. Section \ref{sec:threats} discusses threats to validity.

\section{Motivating Example}\label{sec:motivating-example}

\begin{figure*}[tb]
  \centering
  \incfig{1.0}{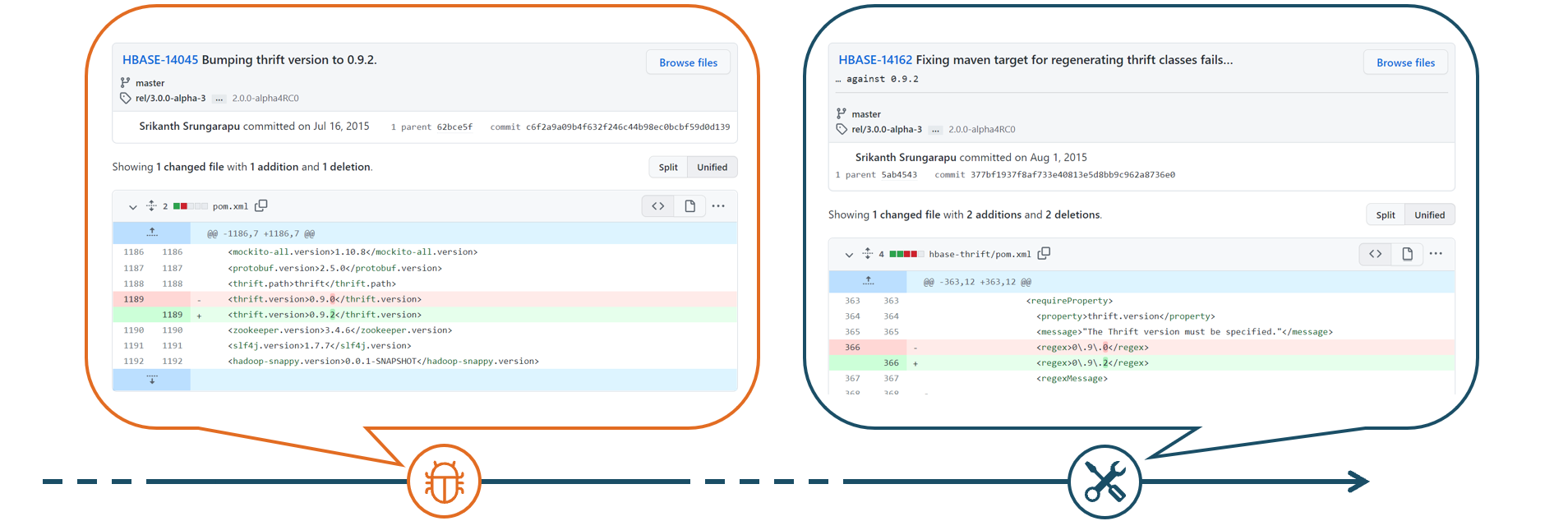}
    \caption{An example of an incomplete change that occurred in real OSS development (HBASE-14045): when the version of a dependent library was upgraded (left), the corresponding modification of the version constraint rule was overlooked (right).}\label{fig:motivating-example-1}
\end{figure*}

\begin{figure*}[tb]
  \centering
  \incfig{1.0}{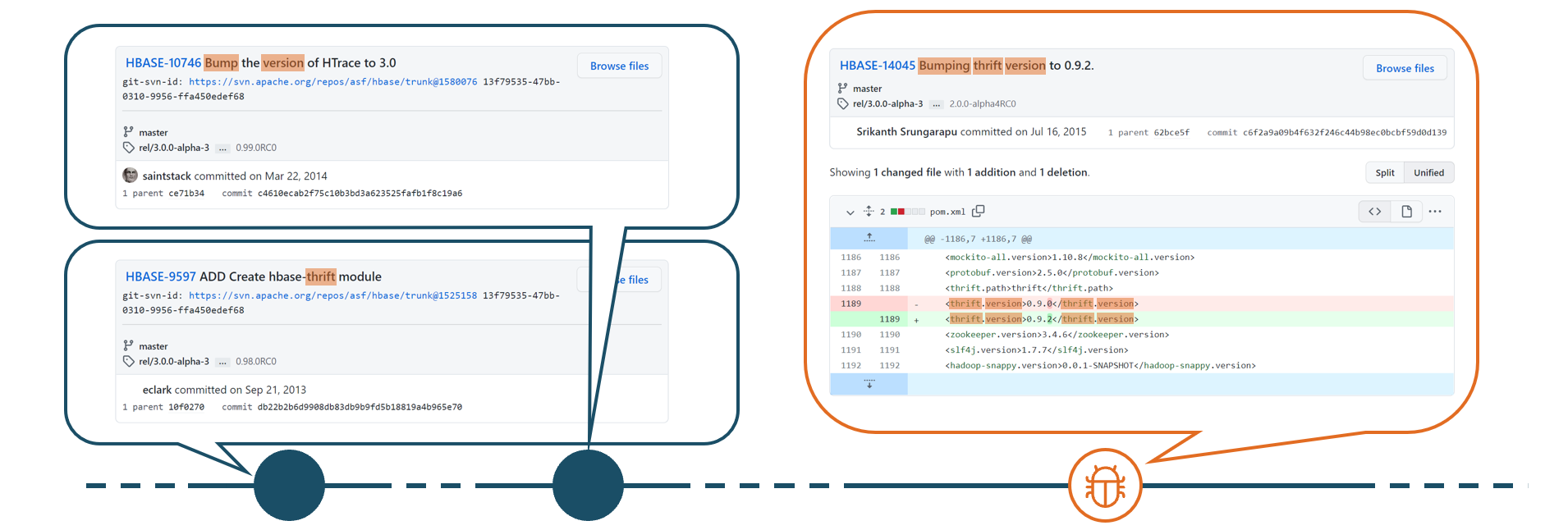}
  \caption{Commits that the proposed method judges to be highly similar to the incomplete change in Fig.~\ref{fig:motivating-example-1}.}\label{fig:motivating-example-2}
\end{figure*}
This section presents an example in which the proposed method is more effective than existing techniques.

Fig.~\ref{fig:motivating-example-1} shows an incomplete change from the dataset that actually occurred in a real OSS project (Apache HBase~\cite{hbase}). The commit shown on the left of the figure\footnote{\url{https://github.com/apache/hbase/commit/c6f2a9a}} modifies a file called \texttt{pom.xml}. This file records project-related information in Maven~\cite{maven}, a project management tool, and the commit in question raises the version of a library on which the project depends. In doing so, the commit fails to update the rule that constrains the version (specified in \texttt{hbase-thrift/pom.xml}), and the missing modification is made in a later commit (shown on the right of the figure)\footnote{\url{https://github.com/apache/hbase/commit/377bf1}}.

In this example, the goal is to recommend \texttt{hbase-thrift/pom.xml} for the change to \texttt{pom.xml}. However, by its very nature, \texttt{pom.xml} has co-change relationships with many different kinds of files, so Evolutionary Coupling alone struggles to recommend the correct element. In practice, with the existing technique~\cite{islam:scam2018}, \texttt{hbase-thrift/pom.xml} is ranked 7th, so the correct element is recommended, but there is still room for improvement in terms of recommendation accuracy.

In contrast, with the proposed method, \texttt{hbase-thrift/pom.xml} rises to 2nd place, which enables change recommendation with higher accuracy. A plausible reason for this result is the introduction of textual information. From the words contained in the commit message and the diff (\textit{bump}, \textit{version}, \textit{thrift}) of the commit in which the incomplete change occurred, it is clear that the version of Thrift~\cite{thrift}, an external library, was upgraded. The proposed method exploits these words to identify, as similar commits, the commit that upgraded the library version (Fig.~\ref{fig:motivating-example-2})\footnote{\url{https://github.com/apache/hbase/commit/c4610e}} and the commit that made changes related to Thrift\footnote{\url{https://github.com/apache/hbase/commit/db22b2}}. Because \texttt{hbase-thrift/pom.xml} is modified in these commits as expected, the correct element can be recommended at a higher rank.

\section{Related Work} \label{sec:relwork}
Ball et al.~\cite{ball:icse1997} and Gall et al.~\cite{gall:icsm1998} pioneered the application of information recorded in version control systems to software analysis, focusing in particular on the co-change relationships that can be extracted from version histories (i.e., which program elements have historically been modified together). The relationships between modules obtained in this manner are known as Evolutionary Coupling~\cite{rolfsnes:saner2016} or Logical Coupling~\cite{gall:icsm1998}. In a study of Apache projects, Oliva et al.~\cite{Oliva2011} found that 91\% of element pairs that are Evolutionarily Coupled have no static dependency between them.

As co-change relationships among program elements become more complex, it becomes harder for developers to identify where a change propagates, which in turn increases the likelihood of defects. Mondal et al.~\cite{Mondal2019} investigated the relationship between Evolutionary Coupling and defects at the method level and found that the number of Evolutionary Couplings established over a given development period is positively correlated with the number of defects. They further showed that methods containing defects have significantly more co-change relationships with other methods than those without. Kirbas et al.~\cite{Kirbas2017} investigated the same relationship at file granularity and confirmed that the number of co-change relationships a file participates in is positively correlated with the number of defects. In contrast, Graves et al.~\cite{Graves2000} built a generalized linear model to predict defects at the module level, where a module is defined as a set of functionally related files, and found that the number of times a module has been co-changed is not a useful predictor of defects compared with other metrics such as the number of revisions to a module. Similarly, Knab et al.~\cite{Knab2006} learned a decision tree that predicts the defect density of modules in Mozilla projects and showed that the number of co-changes is not useful for this prediction. Kirbas et al.~\cite{Kirbas2017} attribute these contradictory results to differences in program element attributes, such as size and number of developers, as well as to differences in context.

Among development-support techniques leveraging Evolutionary Coupling, one major category recommends change locations relevant to the developer's current task. Rose \cite{zimmermann:icse2004} applies association rule mining to version histories and extracts co-change relationships as co-change rules. By identifying elements that have strong co-change relationships with the elements currently being modified, Rose recommends additional change locations.

Although Rose suppresses false positives, it suffers from limited applicability. Rose extracts recommendation candidates only from commits that contain all elements in the query set; if no such commit exists, no recommendation can be produced. To address this limitation, Rolfsnes et al.\ proposed TARMAQ \cite{rolfsnes:saner2016}, which relaxes the constraints on rule extraction. TARMAQ considers all commits whose intersection with the query is maximized, thereby increasing the number of situations in which recommendations can be made. TARMAQ has been reported to outperform Rose in terms of Average Precision.

Rose and TARMAQ rely entirely on past co-change relationships. Thus, elements that have never co-changed with the input elements cannot be recommended. To overcome this limitation, Islam et al.\ proposed Transitive Association Rule Mining \cite{islam:scam2018}. This technique assumes that if A–B and B–C co-change relationships exist, then A–C also has a transitive Evolutionary Coupling. Their method recommends correct elements in more cases than previous co-change-rule-based approaches.

In general-purpose version control systems, changes to a codebase are recorded at commit granularity. Software analyses therefore commonly treat all program elements appearing in a commit alike when extracting Evolutionary Coupling. In practice, however, developers often commit with several distinct intents~\cite{herzig:msr2013}, and treating a commit as a single unit works against the analysis in such cases. To address this, several approaches extend Evolutionary Coupling with additional information and tools. In the Interaction Coupling proposed by Zou et al.~\cite{Zou2007}, a developer declares an ongoing task in the IDE, and operations on program elements are analyzed at the task level rather than the commit level to derive co-change relationships. Robbes et al.~\cite{Robbes2008} instead record the times at which developers modify program elements in the IDE, and weight the strength of Evolutionary Coupling according to the temporal proximity of the modification times of the elements involved.

Poshyvanyk et al.~\cite{poshyvanyk:ese2009} proposed Conceptual Coupling as a measure of how similar the concepts held by each class are. In Conceptual Coupling, the similarity between two methods is measured by the similarity of their comments and identifier names, and these values are then aggregated to the class level. When a developer makes a change, classes that are strongly Conceptually Coupled with the changed location are inferred to be the targets of the change. Mondal et al.~\cite{mondal:csmr2014} proposed incorporating identifier information when measuring the strength of co-change rules: their method computes the diffs between consecutive commits and strengthens co-change rules when similar identifiers appear. Kagdi et al.~\cite{kagdi:wcre2010} combined recommendation based on co-change rules with Conceptual Coupling, but their method simply merges the top $N$ results of the two recommenders to produce the final recommendation.

Hassan et al.~\cite{Hassan2004} examine four heuristics for predicting where a change propagates: DEV, elements previously changed by the same developer; HIS, elements that have previously co-changed with the input in the version history; CUD, elements related through function calls, variable definitions, or similar constructs; and FIL, elements located in the same file. They evaluate these heuristics at the granularity of software entities, including functions, variables, and type declarations. Excluding FIL, whose predictive scope is inherently limited, HIS achieves the highest performance, leading Hassan et al. to recommend exploiting version histories when predicting where changes propagate.

Change requests filed in issue management systems such as Bugzilla~\cite{bugzilla} trigger fixes to the codebase. Because change requests are usually written in natural language such as English, several approaches exploit this textual information to infer change locations. In the approach of Gethers et al.~\cite{Gethers2012}, each program element is represented as a sequence of words by extracting its identifiers and comments. Given a change request, its content is compared with each program element using natural language processing techniques, and the elements with the highest similarity are presented as the predicted change locations. InComIA, proposed by Zanjani et al.~\cite{Zanjani2014}, likewise generates a textual representation of each program element, but additionally incorporates the text of related commit messages and bug reports alongside identifiers and comments. These approaches resemble ours in that they exploit textual information, but they differ from the proposed method in three respects. First, their input is a change request rather than a developer's current change, and a change request presupposes that the change has already been filed and described. Second, they do not exploit information about the changed elements themselves. Third, their similarity is defined between a query and an individual program element, so that the version history serves only as a source of text to be indexed, rather than as the medium through which an unfinished change is located. The proposed method instead retrieves past commits that resemble the work in progress and infers change locations from the elements those commits actually modified.

\v{C}ubran\'{i}\'{c} et al.~\cite{davor:icse2003} built Hipikat, which recommends program elements and issue tickets related to a developers' task. Hipikat constructs a graph consisting of multiple types of nodes, including developers and program elements, based on the textual similarity and the temporal proximity of changes. Given an input, it recommends the entities that are close to it in the Hipikat graph as related. Hipikat is similar to the proposed method in that it exploits similarity along diverse aspects of a change, but it was developed primarily to support new members of a project, and the evaluation of \v{C}ubran\'{i}\'{c} et al. does not consider the identification of missing locations in incomplete changes.

Mondal et al.~\cite{mondal:icpc2021} proposed FLeCCS, a change recommendation technique that applies code clone detection. It searches a project for code fragments that are similar in content to the input change lines and recommends them as change locations. The results show that FLeCCS has room for improvement when combined with Evolutionary Coupling information.

Taken together, prior work divides along two axes. Co-change-rule-based techniques derive recommendations from the version history, but represent each commit only as a set of modified elements, so they inherit the sparseness of the history and do not systematically exploit the text that accompanies a commit. Text-based techniques use such text, but derive it from the source code or from a filed change request, and rank individual elements by their similarity to a query. The proposed method sits at the intersection: it uses the version history as the medium of comparison, as co-change-based techniques do, and it uses the text that accompanies each commit, as text-based techniques do, but it relates the two by treating commits, rather than program elements, as the unit of similarity. Recommendation is then a matter of retrieving the commits most similar to the change in progress and observing which elements those commits modified, which is what distinguishes the proposed method from both classes.

\section{Proposed Method : {\sf GitCF}} \label{sec:method}
The proposed method {\sf GitCF} recommends overlooked change locations by leveraging similarity among commits based on various attributes. It consists of three modules: recommendation based on modified elements (\ref{sec:edit-based}), recommendation based on textual information (\ref{sec:text-based}), and integration of the two recommendation results (\ref{sec:merge}).

\subsection{Recommendation Based on Modified Elements}\label{sec:edit-based}
If a past commit modified program elements similar to those currently being edited, the elements appearing in that commit are likely relevant to the current task. The first module identifies commits in the version history that have high similarity to the current modification in terms of modified elements, and recommends the elements modified in those commits.

To measure similarity between sets of modified elements, this study represents commits as vectors and computes cosine similarity. At recommendation time, we identify the set of program elements $E$ present in the project and represent each past commit as an $N = |E|$-dimensional vector $v$. Using this vector, the similarity $sim$ between the input set $Q$ and a past commit $C$ is defined as in Equation~\eqref{eq:cos_sim}.
\begin{equation}
\label{eq:cos_sim}
  sim(Q, C) = \frac{v(Q) \cdot v(C)}{|v(Q)||v(C)|}
\end{equation}

This study proposes three types of features for the vector $v$ and four similarity measures that combine them.

\subsubsection{Location of Modification (CHG)}\ 

The basic vector representation $v(C)$ for a commit $C$ indicates whether each element was modified:
\begin{equation}
  \label{eq:cos_vec}
  v(C) = ({\mathit chg}_C(e_1), chg_C(e_2), \dots, chg_C(e_N))
\end{equation}
Here, $chg_C(e)$ is 1 if element $e$ was modified in commit $C$, and 0 otherwise.

\subsubsection{Modification Size (LOE)}\

If a program element is heavily modified in a commit, it may be more important for characterizing that commit. We introduce LOE (\textit{Lines Of Edit}) to quantify how extensively an element was modified:
\begin{equation}
  \label{eq:loe}
  loe_{C}(e) = chg_C(e) \cdot \max(1, \max(ins_C(e), del_C(e)))
\end{equation}
Here, $ins_C(e)$ and $del_C(e)$ denote the number of added and deleted lines, respectively. We use the maximum rather than the sum to avoid overemphasizing overlapping additions and deletions. Renames yield LOE = 1.

The LOE-based vector is:
\begin{equation}
  \label{eq:loe_vec}
  v_{loe}(C) = (loe_C(e_1), loe_C(e_2), \dots,loe_C(e_N))
\end{equation}

\subsubsection{Importance Based on Element Frequency (ICF)}\

When a program element within a commit is updated frequently regardless of the actual content of the change, that element does not significantly contribute to characterizing the commit. Therefore, based on insights from prior research on collaborative filtering [14], we introduce ICF ({\it Inversed Commit Frequency}) as a measure of importance that reflects the occurrence frequency of each element. The ICF value for an element $e$ is calculated using Equation (5).
\begin{equation}
  \label{eq:icf}
  icf(e) = \log \frac{|\mathcal{H}|}{\sum_{C \in \mathcal{H}} chg_C(e)}
\end{equation}
where $\mathcal{H}$ is the set of all commits.

We combine ICF with CHG and LOE using the Hadamard product:
\begin{equation}
  \label{eq:icf_vec}
  v_{icf}(C) = (icf(e_1), icf(e_2), \dots, icf(e_N))
\end{equation}

\subsubsection{Score Calculation}\

We select the top-$k$ most similar commits $\mathit{kNN}(Q, \mathcal{H})$ and compute scores:
\begin{equation}
  \label{eq:score}
  score(Q, e) = \frac{1}{t}\sum_{C \in \mathit{kNN}(Q, \mathcal{H})} sim(Q, C) \cdot chg_C(e)
\end{equation}
where $t$ normalizes scores to $[0,1]$. The top-$N$ elements are recommended.

\subsection{Recommendation Based on Textual Information}\label{sec:text-based}
Commit messages ideally contain information about modified locations and intent \cite{yingchen:icse2022}. Likewise, module names, identifiers, and linked issues provide useful context. The second module computes similarity among commits using such textual information.

We extract three types of textual information: commit messages, issue descriptions, and diffs.

Commit messages are used directly (denoted {\bf M}).  
If a commit references an issue ID \cite{sliwerski:sigsoft2005}, we append the issue description ({\bf Mi}).  
Diffs contain noise, so we extract only identifiers using Sun et al.'s regular expressions \cite{sun:ist2017} ({\bf D}).

We also consider combinations {\bf M+D} and {\bf Mi+D}, yielding five settings.

All textual data undergo standard preprocessing (symbol removal, camel-case splitting, stopword removal, stemming) and are vectorized using TF-IDF. Experiments showed that binary TF (0/1) yields the best performance.

Similarity is computed using cosine similarity as in Section~\ref{sec:edit-based}, and scores are computed using Equation~\eqref{eq:score}.

\subsection{Integration of Recommendation Results}\label{sec:merge}
The two modules defined above are each capable of performing change recommendation on their own.
However, in recommender systems, it is also common to combine multiple techniques in order to improve accuracy \cite{burke:um2002}.
We therefore integrate the recommendation results of the two techniques to improve performance.

Methods for combining multiple rankings include approaches that learn how to combine rankings based on training data \cite{cao:icml2007}. For the task at hand, however, constructing training data is difficult. We therefore evaluated the following two methods for integrating recommendation results, neither of which requires training data.

Using these methods, we finally integrate the recommendation results from changed-element-based recommendation and textual-information-based recommendation, and the top $N$ program elements with the highest combined scores are recommended to the developer.

\subsubsection{Reciprocal Rank Fusion (RRF)} \ 

Reciprocal Rank Fusion (RRF), proposed by Cormack et al.~\cite{cormack:sigir2009}, combines rankings without using the scores computed during ranking, and instead relies solely on the ranks at which the targets appear. RRF takes the reciprocal of the rank at which a target appears as its score, and sums the scores from the individual recommendations to obtain the final score. When RRF is used, the score assigned to a program element $e$ is given by Equation~(\ref{eq:rrf}).

\begin{equation}
  \label{eq:rrf}
  \mathit{RRF(e)} = \frac{1}{K + rank_e(e)} + \frac{1}{K + rank_t(e)}
\end{equation}
with $K=60$.

\subsubsection{wCombSUM} \

In CombSUM, the sum of the scores output by each recommendation module is simply taken as the score for integration \cite{lee:sigir1997}. wCombSUM is a variant of CombSUM that introduces weights, and it integrates the score $score_e$ output by the changed-element-based module and the score $score_t$ output by the textual-information-based module with the weight given by the parameter $\alpha$ (Equation~(\ref{eq:wcomb})). The value of $\alpha$ ranges over $[0,1]$. When $\alpha=0.5$, the result is the same as CombSUM without weights.
\begin{equation}
  \label{eq:wcomb}
  \mathit{wCombSUM(e)} = \alpha \cdot \mathit{score_{e}(e)} + (1 - \alpha) \cdot \mathit{score_{t}(e)}
\end{equation}
with $\alpha \in [0,1]$.

\section{Experimental Setup} \label{sec:design}
\begin{table}[tb]
  \footnotesize
  \centering
  \caption{Dataset}\label{tab:dataset}
  \begin{threeparttable}
  \begin{tabular}{lcclcc}
    \hline
    & \#C & \#IC && \#C & \#IC
    \\ \hline 
    Accumulo
    &10,452 &10
    &\hspace{8mm}
    Ignite
    &26,996 &43  \\ 
    Ambari
    &24,584 &15
    &\hspace{8mm}
    Lucene
    &33,403 &19  \\ 
    Calcite
    & 4,280 & 8
    &\hspace{8mm}
    Oozie
    & 2,355 &21  \\ 
    Camel
    &45,786 &17
    &\hspace{8mm}
    Pig
    & 3,694 &40  \\ 
    Cassandra
    &25,363 &34
    &\hspace{8mm}
    Spark
    &27,491 &76  \\ 
    Flink
    &23,310 &36
    &\hspace{8mm}
    Struts
    & 5,950 &21  \\ 
    Hadoop
    &23,670 &21
    &\hspace{8mm}
    Thrift
    & 6,059 &12  \\ 
    Hbase
    &17,803 &71
    &\hspace{8mm}
    Wicket
    &20,951 &32  \\ 
    Hive
    &14,631 &85
    &\hspace{8mm}
    \bf{TOTAL}
    & - &\bf{543}  \\ \hline
    
  \end{tabular}
  \begin{tablenotes}
    \item[] \#C: Number of commits in the version history at dataset creation
    \item[] \#IC: Number of incomplete-change commits
  \end{tablenotes}
  \end{threeparttable}
\end{table}

\subsection{Evaluation Method}
We evaluate using the dataset of real incomplete changes in OSS created by Ishida~\cite{yishida:mthesis2021}. 
Table~\ref{tab:dataset} summarizes the dataset. As defined in Section~\ref{sec:intro}, every incomplete change in this dataset is one in which the omitted modification is supplied in a later commit. Each incomplete-change commit constitutes one task, and performance is measured by whether the forgotten file is recommended.

The number of similar commits $k$ must be determined. We vary $k$ from 10 to 500 in increments of 10 and select the value maximizing MAP. For wCombSUM, we vary $\alpha$ from 0 to 1 in increments of 0.1 and select the value maximizing MAP.

We compare three co-change-rule-based techniques: TARMAQ \cite{rolfsnes:saner2016}, LCExtractor \cite{mori:ipsj2017e}, and TAR \cite{islam:scam2018}. Throughout the evaluation, we refer to TAR as the baseline, because it achieves the highest performance among the three on all metrics except Hit@20, on which it ties with LCExtractor. Following prior work, commits modifying more than 30 elements are excluded as noise. Sorting criteria follow the respective papers.

\begin{itemize}
  \item {\bf TARMAQ}: Sort by Support, then Confidence.
  \item {\bf LCExtractor}: Sort by Confidence, then Support. Maximum rule length is 4.
  \item {\bf TAR}: Sort by Confidence, then Support. Transitive rules have Support = 0.

\end{itemize}

\subsection{Evaluation Metrics}
We use MAP, Recall@N, and Hit@N.

MAP is the mean of AP:
\begin{equation}
  \label{eq:met:ap}
  AP = \sum {\mathit precision(i)} \times \Delta r(i)
\end{equation}

Recall@N measures recall using the top-$N$ recommendations.  
Hit@N measures the proportion of tasks for which at least one correct element appears in the top-$N$ recommendations.

Throughout the evaluation, $N$ is set to 10 and 20. The five metrics reported in Table~\ref{tab:result} are therefore MAP, Recall@10, Recall@20, Hit@10, and Hit@20.

\subsection{Research Questions}
To investigate the performance and practical usefulness of the proposed change recommendation method, we address the following two research questions.

\subsection*{{\sf RQ1}: \textbf{\it Does the proposed method improve recommendation performance over existing techniques?}}

This research question addresses whether the proposed method improves recommendation performance over the baseline. 
Along the way, it identifies the configuration of the proposed method that most effectively exploits changed-element information and textual information, as well as their integration.
It verifies whether MAP, from a comprehensive perspective, and ${Recall}$ and $Hit$, from a practical perspective, are improved. 
To identify the most effective configuration, we define the following three sub-RQs:

\begin{itemize}
  \item {\sf RQ1.1}: {\it Which similarity measure is effective for element-based recommendation?}
  \item {\sf RQ1.2}: {\it Which textual-information setting is effective?}
  \item {\sf RQ1.3}: {\it Is integrating recommendation results effective?}
\end{itemize}

\subsection*{{\sf RQ2}: \textbf{\it How practical is the proposed change recommendation method under real-world constraints?}}

This research question deepens our understanding of the practical conditions under which
the proposed method is effective. Assuming a real-world deployment setting, it examines
the cases in which the proposed method is more effective than existing techniques.

\begin{itemize}
  \item {\sf RQ2.1}: {\it What is the execution time required for recommendation?}

  Unlike existing techniques, the proposed method does not restrict the mining scope.
  Its recommendation time is therefore expected to increase relative to that of existing
  techniques. RQ2.1 examines whether this increase in execution time remains within a
  practically acceptable range.

  \item {\sf RQ2.2}: {\it How does project age relate to performance?}

  The development period of software varies depending on the development setup.
  In particular, a short development period means a short version history available for
  mining, which is likely to affect recommendation performance. RQ2.2 examines whether
  the length of the development period affects recommendation performance.
\end{itemize}

\section{Experimental Results} \label{sec:result}

\subsection{\sf{RQ1}}
Table~\ref{tab:result} summarizes the experimental results for each method.

\subsubsection{\sf{RQ1.1}: Metrics Effective for Recommendation Based on Change Elements} \

As shown in the experimental results, the MAP value for CHG is higher than that of LCExtractor, whereas it is lower than that of TAR. This indicates that CHG slightly outperforms recommendation based on co-change rules that do not consider transitive relations, but does not reach the performance level of TAR.

When ICF is incorporated, the recommendation results improve across most evaluation metrics compared to CHG, which does not consider element occurrence frequency. Thus, the notion of program-element occurrence frequency effectively contributes to change recommendation.

In contrast to the improvement brought by ICF, introducing LOE tends to degrade performance compared to the setting without LOE. A likely reason is that, under situations where change propagation occurs, the number of modified lines does not necessarily reflect the importance of the program elements involved. For example, when a method in a class is renamed, the modification within that class may require only one line, whereas all classes invoking the renamed method must be updated for each invocation. In such cases, the amount of change applied to the origin of the propagation (where the important change actually occurred) becomes smaller than the amount applied to the propagation targets. Therefore, the quantity of change does not reliably correspond to the importance of the element, and the proposed use of line counts does not function effectively.

Consequently, CHG+ICF is adopted as the representation for the module based on change elements.

\begin{table}[tb]
\setlength{\tabcolsep}{5pt}
\renewcommand{\arraystretch}{1.15}
\footnotesize
\centering
\caption{Experimental results}\label{tab:result}
\begin{threeparttable}
\begin{tabular}{p{1.1cm}|c|cc|cc|c}
\hline
&& \multicolumn{2}{c|}{$Recall$} & \multicolumn{2}{c|}{$Hit$} & \multirow{2}{*}{MAP} \\ \cline{3-6}
&& $@10$ & $@20$ & $@10$ & $@20$ & \\ \hline \hline
\multirow{3}{*}{\begin{tabular}{@{}l@{}}Baseline\end{tabular}}
&TARMAQ\cite{rolfsnes:saner2016}
& 0.104&0.124 & 0.147&0.168 & 0.054 \\ \cline{2-7}
&LCExtractor\cite{mori:ipsj2017e}
& 0.220&0.291 & 0.320&\bf{0.411} & 0.123 \\ \cline{2-7}
&TAR\cite{islam:scam2018}
& \bf{0.228}&\bf{0.293} & 0.322&\bf{0.411} & \bf{0.135} \\ \hline
\multirow{7}{*}{\begin{tabular}{@{}l@{}}Change\\Elements\end{tabular}}
&\begin{tabular}{c}CHG\\($k=50$)\end{tabular}
& 0.230&0.312& 0.337&0.435& 0.129 \\ \cline{2-7}
&\begin{tabular}{c}LOE\\($k=90$)\end{tabular}
& 0.204&0.265& 0.309&0.390& 0.129 \\ \cline{2-7}
&\begin{tabular}{c}CHG+ICF\\($k=120$)\end{tabular}
& \bf{0.230}&\bf{0.325}& \bf{0.344}&\bf{0.462}& \bf{0.138} \\ \cline{2-7}
&\begin{tabular}{c}LOE+ICF\\($k=400$)\end{tabular}
& 0.208&0.281& 0.317&0.403& 0.134 \\ \hline
\multirow{9}{*}{\begin{tabular}{@{}l@{}}Textual\\Info.\end{tabular}}
&\begin{tabular}{c}M\\($k=130$)\end{tabular}
& 0.160&0.225& 0.267&0.343& 0.092 \\ \cline{2-7}
&\begin{tabular}{c}Mi\\($k=160$)\end{tabular}
& 0.180&0.249& 0.291&0.378& 0.096 \\ \cline{2-7}
&\begin{tabular}{c}D\\($k=50$)\end{tabular}
& 0.235&0.318& 0.339&0.427& 0.135 \\ \cline{2-7}
&\begin{tabular}{c}M+D\\($k=30$)\end{tabular}
& \bf{0.253}&\bf{0.327}& \bf{0.372}&\bf{0.449}& \bf{0.142} \\ \cline{2-7}
&\begin{tabular}{c}Mi+D\\($k=30$)\end{tabular}
& 0.249&0.324& 0.363&0.448& 0.139 \\ \hline
\multirow{3}{*}{Integration}
&RRF
& \bf{0.285}&\bf{0.362}& \bf{0.416}&\bf{0.490}& 0.153 \\ \cline{2-7}
&\begin{tabular}{c}wCombSUM\\($\alpha=0.3$)\end{tabular}
& 0.278&0.358& 0.403&0.484& \bf{0.160} \\ \hline
\end{tabular}
\end{threeparttable}
\end{table}

\subsubsection{\sf{RQ1.2}: Effective Settings for Recommendation Using Textual Information} \ 

Among the proposed methods, recommendation using only commit messages (M) exhibits lower performance than both the baseline and the other methods. Although incorporating issue information (Mi) slightly improves performance, it still does not outperform other approaches.

On the other hand, the settings that leverage change diffs (D, M+D, and Mi+D) achieve higher performance across all metrics than the settings without diffs. Notably, M+D achieves the highest MAP among all non-integrated methods. This demonstrates that exploiting change diffs is particularly effective for change recommendation.

Comparing M+D and Mi+D, the latter performs worse. This suggests that issue information is somewhat effective when combined with commit messages, but its benefit diminishes when used together with change diffs. Further analysis of the performance degradation in Mi+D and exploration of effective ways to utilize issue information remain future work.

Based on these results, M+D is adopted as the textual-information-based recommendation method.

\subsubsection{\sf{RQ1.3}: Effectiveness of Integrating Recommendation Results} \ 

From the experiments in {\sf RQ1.1} and {\sf RQ1.2}, CHG+ICF ($k=120$) and M+D ($k=30$) were identified as the best settings among our module configurations. Here, we examine whether integrating their recommendation results can further improve performance.

The experimental results show that both RRF and wCombSUM improve all evaluation metrics compared to non-integrated settings. In particular, $Hit@10$ exceeds 0.4, indicating that under conditions where incomplete changes are known to occur, useful recommendations appear within the top 10 results in 40\% of cases. Moreover, Wilcoxon signed-rank tests on 
AP show significant improvement over the baseline (TAR) for both integration methods.

Comparing RRF and wCombSUM, wCombSUM achieves a higher 
MAP. Therefore, wCombSUM is adopted as the final proposed method. However, RRF also yields significant improvement over the baseline and has the advantage of requiring no parameters such as 
$\alpha$, making it a viable option for future extensions.

\begin{screen}
\textbf{\sf{Answer to RQ1}:} Among the methods examined, the combination of ``CHG+ICF'', ``M+D'', and ``wCombSUM'' achieved the highest recommendation performance, with a statistically significant improvement in $AP$ compared to the baseline.
\end{screen}

\subsection{\sf{RQ2}}

\subsubsection{\sf{RQ2.1}: Execution Time Required for Recommendation} \ 

\begin{figure}[tb]
\centering
\incfig{1.0}{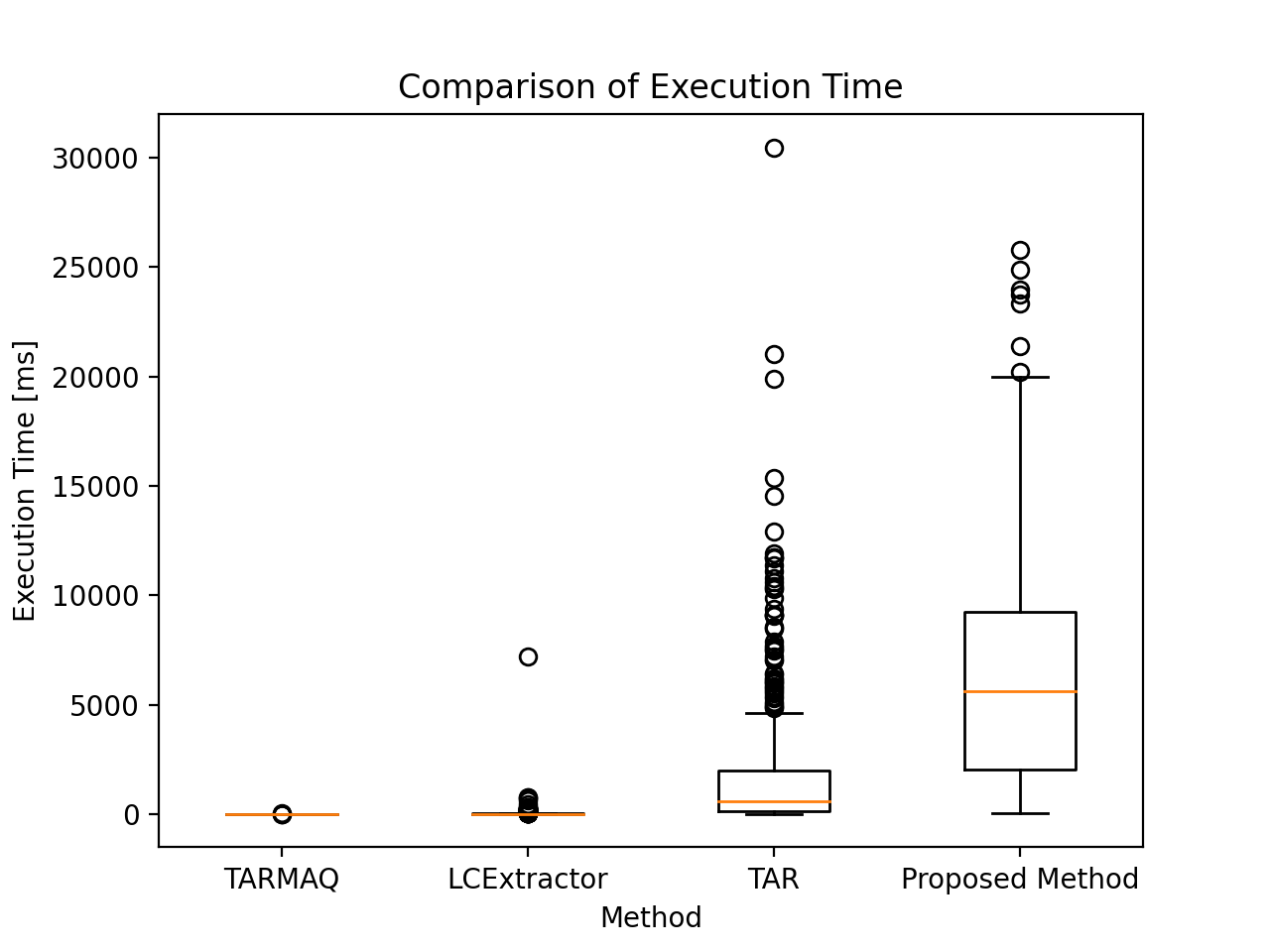}
\caption{Comparison of execution time}
\label{fig:exec-time}
\end{figure}

We compare the execution time of each change recommendation method. Figure~\ref{fig:exec-time} plots, in milliseconds, the time required for each method to perform a single recommendation task. Here, the proposed method refers to wCombSUM.

Compared to the baseline, the proposed method tends to require the longest execution time. Because the proposed method incorporates textual information, it searches all commits in the revision history regardless of input. As a result, it incurs higher computational cost than rule-based methods that restrict the mining scope.

Nevertheless, even the proposed method completes all recommendation tasks within 30 seconds. Although this may be somewhat long for scenarios requiring immediate and frequent recommendations within an IDE, it remains acceptable for checking missing changes at specific moments, such as when developers commit or when changes are pushed to a remote repository.

\subsubsection{\sf{RQ2.2}: Relationship Between Development Period Length and Recommendation Performance}\

\begin{figure*}[tb]
\centering
\incfig{0.8}{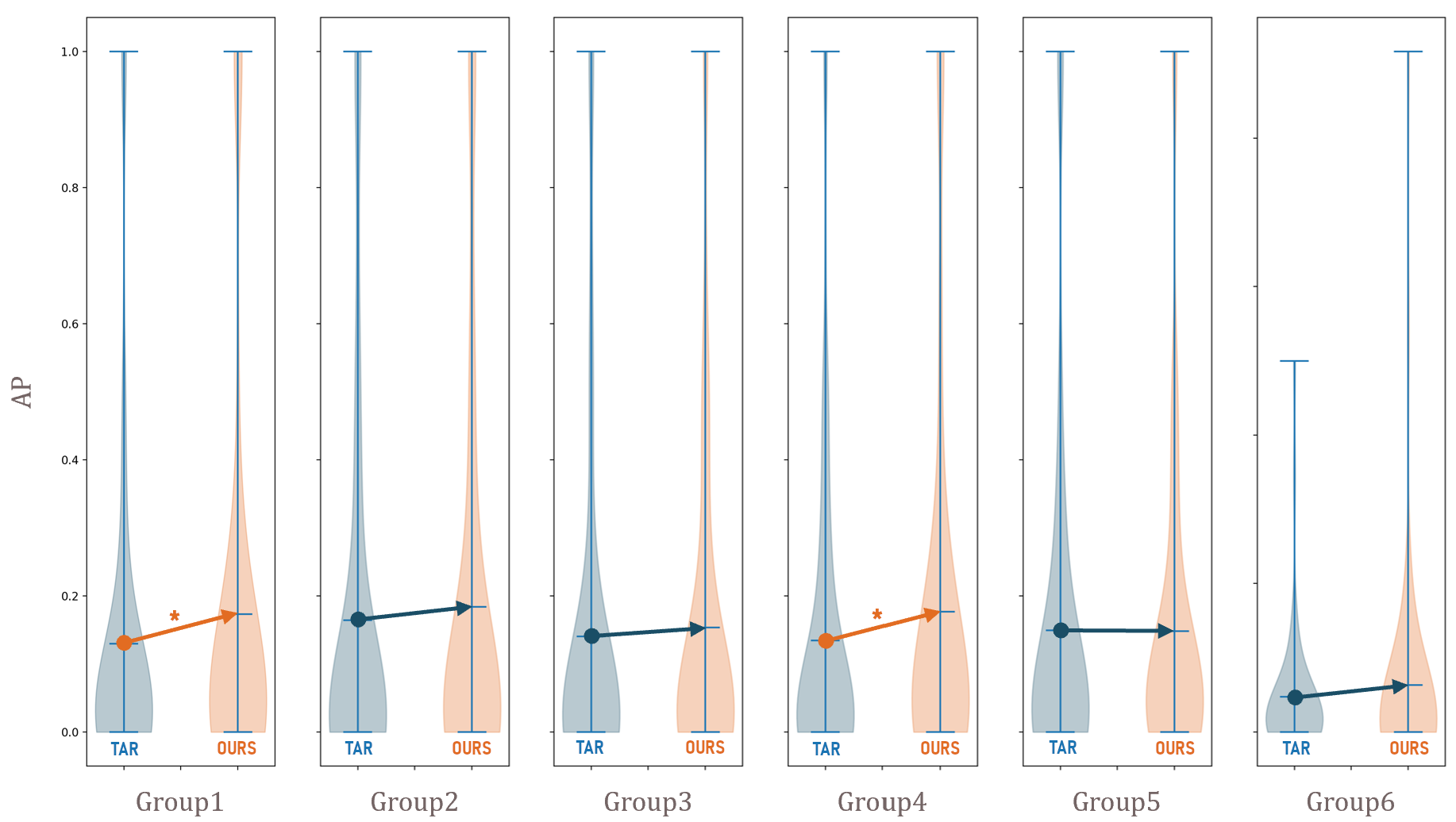}
\caption{Relationship between development period and recommendation performance}
\label{fig:ap-devcommits}
\vspace{-1em}
\end{figure*}

In the early stages of development, the revision history contains few commits, so useful Evolutionary Coupling information cannot be obtained from the past. This is the well-known cold-start problem. As a result, co-change-rule-based methods are limited in the rules available to them and tend to exhibit lower recommendation performance.
In contrast, the proposed method leverages textual information, which is less affected by the length of the revision history. The remainder of this subsection examines how the length of the development period relates to recommendation performance, in order to assess the effectiveness of the proposed method during early development.

In this experiment, to examine the effect of the development period, we sort the recommendation tasks in the evaluation dataset in ascending order by the number of commits that precede each incomplete-change commit, and divide them into groups of 100 tasks.
Group 1, for example, consists mainly of incomplete-change commits that occurred when the total number of commits was approximately 5{,}000, so it collects the cases from the early stages of development across the projects.
To examine the effect, we measure the $AP$ of both the baseline (TAR) and the proposed method for each of these groups.
Figure~\ref{fig:ap-devcommits} is a violin plot that shows the task-level $AP$ of each group.
The arrows in the plot connect the medians. The orange color indicates the pairs for which a significant difference was observed in the Mann-Whitney U test at $p < 0.05$, whereas the blue color indicates the pairs for which no significant difference was confirmed.
The results show that there is little difference in performance in Group 5, which corresponds to longer development periods, whereas in Group 1, which corresponds to shorter development periods, the proposed method yields a comparatively large significant improvement in $AP$.
Indeed, in Group 1 there is a statistically significant difference between the baseline (TAR) and the proposed method.

These results indicate that the proposed method is less affected by revision-history length and remains effective even in early development. Furthermore, because fewer commits exist in early stages, the long execution time noted in the previous subsection is less likely to become problematic. Thus, the proposed method is particularly effective when applied early in a project.

\begin{screen}
\textbf{\sf{Answer to RQ2}:} Although the proposed method requires more execution time than the baseline, the time remains within an acceptable range. Moreover, unlike existing methods, the proposed method is effective even during the early stages of a project.
\end{screen}

\section{Threats to Validity}\label{sec:threats}
\subsection{Internal Validity}
A potential threat to internal validity concerns the implementation of the proposed method. Although both the proposed techniques and the baseline techniques were re-implemented with care, the implementations may contain unintended defects that could influence the results. Such defects, if present, are not necessarily observable through the experimental procedures used in this study.

\subsection{External Validity}
A threat to external validity arises from the characteristics of the dataset used in the experiments. All projects included in the dataset are Apache projects, and differences in development culture or project conventions may limit the applicability of the findings to other ecosystems. Furthermore, the granularity of the program elements targeted for recommendation was set to the file level. Applying the method at a finer granularity, such as the method level, may lead to different outcomes, and the extent to which the findings generalize across granularities remains uncertain.

\subsection{Construct Validity}
This study employs regular expressions to locate identifiers in change diffs for the purpose of extracting textual information. A threat to construct validity arises because the regular expressions may extract content that is not an identifier, and because relevant information may exist in portions of the diff from which no identifiers are extracted. In addition, the amount of change is represented using the larger of the number of added lines or deleted lines. This choice introduces a construct validity threat, as the line is used as the unit of measurement rather than a finer-grained unit such as the token, which may capture different aspects of change magnitude.

\section{Conclusion}
This paper presents {\sf GitCF}, a change recommendation method that leverages commit attributes and similarities between commits, with a focus on contextual information surrounding changes. We applied GitCF to a dataset of incomplete changes observed in OSS projects and compared it with three baseline techniques. Across the evaluation metrics used in this study, GitCF yielded higher scores than the baselines. For example, in the case of $Hit@10$
, which has been regarded as practically relevant in prior work~\cite{rolfsnes:saner2016}, GitCF achieved a score of 0.403 compared with 0.322 for the baseline, corresponding to an absolute difference of 0.081 and a relative difference of 25.2\%. Among the commit attributes examined, three attributes from previous commits ('program-element occurrence frequency', 'commit messages', and 'change diffs') contributed to the observed improvements.

GitCF requires longer execution time than the baseline techniques. However, because recommendations do not need to be generated continuously, the additional cost may be acceptable depending on the usage context. In addition, an analysis of development period length and recommendation performance suggests that GitCF may be particularly beneficial when the revision history is short.

Future work includes examining recommendation methods that incorporate characteristics of source code more extensively. For example, when using change diffs, this study removed only general stopwords; removing programming-language-specific frequent tokens (e.g., \texttt{get}, \texttt{build}) may influence performance and warrants further investigation.